\documentclass[aps,prb,twocolumn,superscriptaddress,showpacs,preprintnumbers,amsmath,amssymb]{revtex4-2}
\usepackage{graphicx}
\usepackage{bm}
\usepackage{enumerate}
\usepackage{mathrsfs}
\usepackage{float}
\usepackage{url}
\usepackage{makecell}
\usepackage{natbib}
\usepackage{multirow}
\usepackage[titletoc]{appendix}
\usepackage{color}

\usepackage{bookmark}
\usepackage{natbib}
\usepackage{hyperref}
\hypersetup{nolinks=false,
urlcolor=cyan, 
citecolor=violet, 
anchorcolor=yellow,
linkcolor=blue, 
citecolor=red,
colorlinks=true}
\usepackage{ragged2e}
\usepackage{multirow}
\usepackage{eucal}

\begin{document}
\title{Wave functions in the Critical Phase: a Planar \textit{Sierpi\'{n}ski} Fractal Lattice}

\author{Qi Yao\href{https://orcid.org/0000-0002-0522-2820}{\includegraphics[scale=0.06]{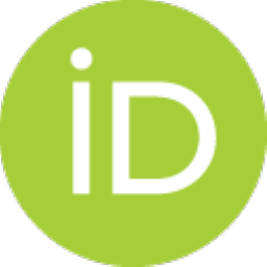}}}
\thanks{These authors contributed equally to this work.}
\affiliation{Quantum Science Center of Guangdong-Hong Kong-Macao Greater Bay Area (Guangdong), Shenzhen 518045, China}
\affiliation{Institute of Quantum Precision Measurement, State Key Laboratory of Radio Frequency Heterogeneous Integration, College of Physics and Optoelectronic Engineering, Shenzhen University, Shenzhen 518060, China}

\author{Xiaotian Yang\href{https://orcid.org/0000-0002-5780-6166}{\includegraphics[scale=0.06]{ORCIDiD.pdf}}}
\thanks{These authors contributed equally to this work.}
\affiliation{Key Laboratory of Artificial Micro- and Nano-structures of Ministry of Education and School of Physics and Technology, Wuhan University, Wuhan, Hubei 430072, China}

\author{Askar A. Iliasov\href{https://orcid.org/0000-0003-2409-7292}{\includegraphics[scale=0.06]{ORCIDiD.pdf}}}
\affiliation{Department of Physics, University of Zurich, Winterthurerstrasse 190, 8057 Zurich, Switzerland}

\author{Mikhail I. Katsnelson\href{https://orcid.org/0000-0001-5165-7553}{\includegraphics[scale=0.06]{ORCIDiD.pdf}}}
\affiliation{Institute for Molecules and Materials, Radboud University, Heijendaalseweg 135, 6525 AJ Nijmegen, The Netherlands}

\author{Shengjun Yuan\href{https://orcid.org/0000-0001-6208-1502}{\includegraphics[scale=0.06]{ORCIDiD.pdf}}}
\email[Corresponding author: ]{s.yuan@whu.edu.cn}
\affiliation{Key Laboratory of Artificial Micro- and Nano-structures of Ministry of Education and School of Physics and Technology, Wuhan University, Wuhan, Hubei 430072, China}
\affiliation{Wuhan Institute of Quantum Technology, Wuhan 430206, China}

\date{\today}

\begin{abstract}
   Electronic states play a crucial role in many quantum systems of moire superlattices, quasicrystals, and fractals. As recently reported in \textit{Sierpi\'{n}ski} lattices [Phys. Rev. B 107, 115424 (2023)], the critical states are revealed by the energy level-correlation spectra, which are caused by the interplay between aperiodicity and determined self-similarity characters. In the case of the \textit{Sierpi\'{n}ski Carpet}, our results further demonstrate that there is some degree of spatial overlap between these electronic states. These states could be strongly affected by its `seed lattice' of the $generator$, and slightly modulated by the dilation pattern and the geometrical self-similarity level. These electronic states are multifractal by scaling the $q$-order inverse participation ratio or fractal dimension, which correlates with the subdiffusion behavior. In the $gene$ pattern, the averaged state-based multifractal dimension of second-order would increase as its \textit{Hausdoff dimension} increases. Our findings could potentially contribute to understanding quantum transports and single-particle quantum dynamics in fractals.
\end{abstract}

\maketitle
\section{Introduction}
Translational invariance of atomic arrangement in crystals leads to the band theory based on the Bloch theorem; Bloch character of electronic wave functions describes enormous amount of properties of crystalline solids \cite{kittel2021introduction,ashcroft1976,vonsovsky1989,phillips2012advanced,el2020advanced}. This symmetry is absent in randomly disordered systems \cite{mott1971,ziman1979,lifshitz1988} as well as in incommensurate (quasiperiodic) systems such as 1D Fibonacci chain~\cite{Shechtman1984Mettallic}, 2D Penrose tiling~\cite{Tsunetsugu1991ele1, Tsunetsugu1991ele2}, Ammann-Beenker lattice~\cite{beenker1982algebraic,grunbaum1987tilings}), and hierarchical tiling of \textit{Sierpi\'{n}ski} lattices~\cite{Mandelbrot1982fractal,feder2013fractals,nakayama2003fractal} and Koch fractals~\cite{Biswas2023Complete}.
In all these cases the tools dramatically different from the band theory are required. In particular, the concept of Anderson localization \cite{Anderson1958Absence} is necessary to understand the properties of disordered systems \cite{mott1971,ziman1979,lifshitz1988}. After many years of efforts, we have well developed mathematical tools to describe this and related phenomena \cite{mirlin2008}, and some of these tools will be named below. The cases of regular but not translationally invariant systems such as quasicrystals and fractals are much less studied, and we are still far from more or less complete understanding of their electron properties.

To better study these aperiodic structures, many researchers have developed various alternative theoretical methods, including renormalization group technique~\cite{Domany1983Solutions,Niu1986Renormal,Chakrabarti1991Renormal,Yan1992Renormal,You1992Local,Sil1993Extended,Chakrabarti1994Renormalization,Chakrabarti1996Exact,Biswas2023Designer}, transfer matrix~\cite{Macia1996Physical,Eilmes1998two,Zbigniew1999Physical,Jagannathan2021Fibonacci}, level-spectra statistics~\cite{Dyson1962I} from random matrices theory~\cite{mehta1991Random,efetov1999supersymmetry},
one-parameter scaling based on studies of
correlated length for coherent structures at different size ~\cite{Cafiero1991Generalized,Nakayama1994Dynamical} or localized length~\cite{Mirlin2000Statistics} in Anderson model~\cite{Anderson1958Absence,Zharekeshev1995Scaling}, the state-based multifractality scaling~\cite{Jagannathan2021Fibonacci}, studies of transport properties~\cite{Amin2022Multifractal,Barbosa2022Turbulence,Yang2022Electronic} etc.

Following the above timeline, two key objects, namely energy spectra~\cite{Dyson1962I,mehta1991Random,efetov1999supersymmetry} and wave functions in real space~\cite{Cafiero1991Generalized,Nakayama1994Dynamical}, come in view, which can be exploited to analyze electronic~\cite{Kohmoto1986Electronic,Mace2017Critical,Zhou2023Time} and phonon~\cite{Quilichini1997Phonon} systems. For instance, the wave functions belong to one of three types: the characteristically (Bloch) extended state~\cite{phillips2012advanced} in periodicity-translation crystals; the typically localized state~\cite{Anderson1958Absence} in disorder-induced systems by impurities, defects, etc; or the state behaves in between and remains critical in several quasicrystals likely Fibonacci chain~\cite{Kohmoto1987Critical}, Thue-Morse lattice~\cite{Chakrabarti1995Role} or some Penrose tiling~\cite{Sutherland1986Self}. In general, these could be distinguished as follows: whereas the localized wave functions have exponentially localized envelope with $|\psi(r)| \sim \exp(-\alpha r^\beta)$, where $\alpha$ and $\beta$ are the spatial parameters, the decay of the critical states following the power-law form with $|\psi(r)| \sim |r|^{-\alpha}$ (or more complicated envelope, e.g., the localized edge states in ring shape~\cite{Kohmoto1986Electronic} and critical states with the SKK form in Penrose lattices~\cite{Mace2017Critical}), respectively. Note that \textit{in the absence of disorder}, lattice frustration possibly induces the fragmented states (i.e., critical states) in quasicrystals and fractals, these states also become more complex with various long-range orders.

\textit{Sierpi\'{n}ski Carpet} $SC(n,m,g^*)$ as a class of fractals~\cite{Mandelbrot1982fractal,George2006Nanoassembly,Fan2014Fractal,Shang2015Assembling,Liu2021Sierpi}, where $(n,m)$ is the parameters of the $generator(n,m)$ and the geometrical hierarchy level $g^*$, it resembles a periodic square lattice in 2D, whose order of ramification is infinite. Experimentally, these fractal objects can be accessed by arranging waveguide tube~\cite{Yang2020Photonic} and electric circuit~\cite{Song2020Realization} or printing acoustic lattices~\cite{Li2023Fractality} in desired fractal shapes~\cite{Mandelbrot1982fractal,Kempkes2018Design}. In Ref.~\cite{Yao2023Energy}, when a single electron roams upon the $SC(n,m,g^*)$ lattices, we have found that all electronic states reside in the critical phase, which isolates that near mobility edge where Anderson transition occurs~\cite{Zharekeshev1995Scaling}. This trait affects their observable properties of quantum conductance~\cite{van2016Quantum} and QC-based box-counting dimension, plasmas~\cite{Westerhout2018Plasmon}, and Hall conductance~\cite{Iliasov2020Hall}, among others.

To obtain more insight into fractal lattices, we would focus on probing the spatial envelope of electronic states, rather than the nearest energy-correlation spectra. And two aspects are studied: (i) the state envelope in these $SC(n,m,g^*)$ lattices at various energy bands; (ii) how other factors, such as the $generator(n,m)$, dilation pattern, fractal dimension $\mathcal{D}_H$, and geometrical hierarchy level $g^*$, affect the critical states (CSs)? It is potentially crucial for further investigating the disorder-induced localization~\cite{Yao2022talk} or many-body correlation effect~\cite{iliasov2024strong} in \textit{Sierpi\'{n}ski} fractal lattices.

The rest of the paper is organized as follows. In Sec.~\ref{SecII:LMM}, we introduce the single-electron gas model on the $SC(n,m,g^*)$ lattices and the state-based multifractality analysis. Our results are presented in Sec.~\ref{SecIII:Result}. Under three lattice-dilation patterns with $SC(n,m,3)$ lattices, the electronic states are sketched by their density profile in Sec.~\ref{SecIII:cases}, and whose multifractal properties are shown in Sec.~\ref{SecIII:Multifrac}. A conclusion is reached in Sec.~\ref{SecIV:Con.}.
\section{lattice, Model, and Method}\label{SecII:LMM}

\emph{Fractal lattices and model}.\textbf{--}
First, we retrospect the fractal $SC(n,m,g^*)$ lattices that we denoted previously, see Fig. 1 in Ref.~\cite{Yao2023Energy}. Using the $generaotr(n,m)$ and two illustrations of the $self$ and $gene$ patterns ($M_{se}$ and $M_{ge}$), various $SC(n,m,g^*)$ lattices are dilated. Here, these two patterns are our main interest, adding the $vari$ pattern as a comparison, which is a variation of the $self$ pattern. Second, a noninteracting electron gas is confined in the $SC(n,m,g^*)$ lattices, which is modeled by
\begin{eqnarray}\label{eq1}
H=-t\sum_{\langle i,j\rangle}(\mathbf{c}_i^\dagger \mathbf{c}_j+\mathbf{c}_j^\dagger \mathbf{c}_i)+V\sum_{i}f(i) \mathbf{c}_i^\dagger \mathbf{c}_i.
\end{eqnarray}
The first term describes a single electron hops between the nearest neighbor site pair $\langle i,j\rangle$. Setting the strength $t$ as the energy unit. The on-site potential with strength of $V$ includes in the second term, which is tailored locally by the function $f(i)$ in Anderson~\cite{Anderson1958Absence}, Harper~\cite{Harper1955Single}, Aubry or Aubry-Andr\'{e}~\cite{Aubry1980analyticity}, and Aubry-Andr\'{e}-Harper model~\cite{Aubry1980analyticity,Harper1955General}. Third, for consistency with Ref.~\cite{Yao2023Energy}, and only taking the lattice topology effect into account, we set up $V=0$. The results of the level-spectra statistics indicate that the electronic states might be intermediately critical (in other words, they always partially occupy the entire lattice, or the tail of the level-correlation spectra follows the power-law trait). We further use the new tool to quantify these electronic states.

\emph{State-based multifractality analysis}.\textbf{--}
For arbitrary state $\psi$, measuring its spatial extension in lattice, as a convenient tactic, can reveal some inherent traits, which is exploited in studying the localization problem with several concepts such as localization length~\cite{Soukoulis1984Fractal,de1989Localization}, structural entropy~\cite{Varga1999One,Pipek1992Universal}, participation ratio~\cite{Pipek1992Universal}, and multifractality~\cite{Jagannathan2021Fibonacci}. We adopt the $2q$-norm multifractality~\cite{Jagannathan2021Fibonacci} by the given formula,
\begin{eqnarray}
\chi_q(\psi, {\color{blue}\Omega}) &=& \frac{\sum_{i \in \mathcal{R}}\left|\psi(i)\right|^{2 q}}{\left(\sum_{i \in \mathcal{R}}\left|\psi(i)\right|^2\right)^q}, \label{eq2}
\end{eqnarray}
with $q$-order inverse participation ratio (IPR) $\chi_q(\psi, \Omega)$, where $\Omega$ is the number of site counted in region $\mathcal{R}$.

To understand Eq.~\ref{eq2} well, we take three examples: i) one of wave function is evenly extended in the spatial lattice, i.e. $\psi(r)=$const, then we have $\chi_q(\psi, \Omega)=\Omega^{1-q}$ with the lattice size $\Omega$; ii) assuming $\psi(r)$ have the power-law-decay envelope with an exponent $\alpha$, instead
\begin{eqnarray}\label{eq5}
\chi_q(\psi, \Omega) \simeq\left\{\begin{array}{ll}
\Omega^{-2(q-1)} & (0 \leq \alpha<1 / q) \\
\Omega^{-2 q(1-\alpha)} & (1 / q \leq \alpha<1), \text { for } q>1. \\
\Omega^{0} & (1 \leq \alpha);
\end{array}\right.
\end{eqnarray}
iii) another one has the envelope of exponential decaying and an oscillation tail, namely the localized state. If the third state continues to degenerate highly and only occupies several sites, we refer to it as the confined state. One might imagine the single-site-occupied state, obviously having $\chi_q(\psi, \mathcal{R})=1$. Note that $\chi_q(\psi, \mathcal{R})$ depends obviously on the region $\Omega(\mathcal{R})$ one computes. Because different states are associated with lattices of various sizes, it becomes tricky to compare these electronic states directly.

However, using $\mathcal{D}_q^\psi(\psi)$ could avoid this issue.
Since the quantity $\chi_q(\psi, \mathcal{R})$ scales exponentially as $\mathcal{R}^{-(q-1)\mathcal{D}_q^\psi(\psi)}$, which is linked the $q$-th fractal dimension~\cite{Wang2020One},
\begin{eqnarray}
\mathcal{D}_q^\psi(\psi) &=& \lim _{\mathcal{R} \rightarrow \infty} \frac{-1}{q-1} \frac{\log \chi_q\left(\psi(i), \mathcal{R}\right)}{\log \Omega\left(\mathcal{R}\right)}.\label{eq3}
\end{eqnarray}
In general, the second-order quantities of $q=2$ demarcate the extended (localized) state with $\chi_2(\psi)=\Omega ^{-1}$ and $\mathcal{D}_2^\psi(\psi)=1$ ($\chi_2(\psi) \simeq 1$ and $\mathcal{D}_2^\psi(\psi)\simeq 0$). There is a special situation where our observed values are just in between, leading the system to a critical state. It can happen in two scenarios. One is near the mobility edge when Anderson transition occurs~\cite{Janssen1981Statistics,Janssen1994multifracal}, The other scenario is found in quasicrystals~\cite{Jagannathan2021Fibonacci}, where geometric frustration plays a role. In both cases, the electronic states are spatially fragmented differently.

Our work mainly explores the CSs within fractals. It is interesting to note that when two CSs are close in characteristics, we can spot the differences between them by measuring the IPR $\chi_q(\psi, \mathcal{R})$ and the state-based fractal dimension $\mathcal{D}_q^\psi(\psi)$,  especially when using a higher value of $q$. This aspect becomes crucial when we observe the vast clustering of CSs in the spectra of $\mathcal{D}_q^\psi(\psi)$.

\begin{figure*}[!htp]
 \flushleft
       \includegraphics[width=6.72cm,height=6.15cm]{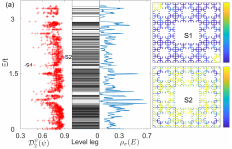}\hskip -0.2pt
       \includegraphics[width=6.72cm,height=6.15cm]{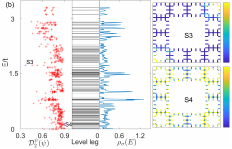}\hskip -0.2pt
       \includegraphics[width=4.45cm,height=6.15cm]{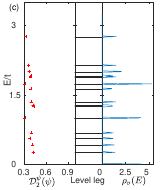}
\caption{(Color online) Schematic illustration of the $SC(n,m,g^*)$ fractal lattices by a paradigmatic $generator(4,2)$ under the \textit{self} (a), \textit{gene} (b), and auxiliary $vari$ pattern (c), respectively. Geometrical hierarchy level $g^*=3$. Energy spectra are symmetric about $E=0$; hence, the upper half panel is considered, i.e., $0\leq E\leq 4t$. At left subpanel in (a), (b), and (c), fractal dimension $\mathcal{D}_2^\psi(\psi)$ (red cross dot) of wave function measures its degree of spatial extension in the whole lattice; level leg $E$ (black line in middle subpanel) and DoS $\rho_\sigma(E)$ (blue curve) with a blurred energy width $\sigma=0.0056t$~\cite{note_DoS} in right subpanel. Two extreme (minimum and maximum) cases of $\mathcal{D}_2^\psi(\psi)$ give four states, which are marked in blue with $S1$ and $S3$ ($S2$ and $S4$) in the upper (bottom) subfigure. The scaled probability density $Ar|\psi_{E_n}(i)|^2$ maps into a colorbar region from 0 to $Ar$.}\label{Fig1:Sketch}
\end{figure*}

\section{Results and Discussion}~\label{SecIII:Result}
In this section, we would study the critical states from different perspectives. These include: i) examining their density profile, which helps us understand how the wave function occupies the entire fractal lattice, akin to analyzing the IPR; ii) investigating the multifractality scaling (fractal dimension), which is related to the subdiffusion behaviors, and could be observed by the dynamical evolutions in larger fractals.

\subsection{The density profile on three $SC(n,m,3)$ lattices}~\label{SecIII:cases}
To highlight how the CSs in fractal lattices are distinct from the other two types of states—extended and localized—we start by examining their spatial density profiles. We demonstrate this using three examples of $SC(4,2,g^*=3)$ lattices. These lattices are designed sequentially following specific patterns: the $self$ pattern with a matrix $M_{se}=[1\;1\;1; 1\;0\;1; 1\;1\;1]$, the $gene$ pattern with $M_{ge}(1)=[1\;1\;1\;1;  1\;0\;0\;1; 1\;0\;0\;1;  1\;1\;1\;1]$, and the $vari$ pattern, a variation of the `self' pattern, with $M_{va}=[1\;0\;0; 0\;0\;1; 1\;1\;0]$. Note that in the $M_{ge}$ pattern, 1 means that the seed lattice of the $generator(n,m)$ is filled, otherwise it is not filled. The other two patterns are similar. The first two patterns ($M_{se}$ and $M_{ge}$) are our primary focus, and the third one serves as a variant to emphasize the potential enhancement effect of local energy clusters (refer to Fig. S1 in our previous work~\cite{Yao2023Energy}).

The level leg in Fig.~\ref{Fig1:Sketch} demonstrates that several critical states form a subband cluster, influenced by specific parameters such as the $generator(n,m)$, the geometrical hierarchy level $g^*$, and the dilation patterns represented by matrices $M_{se}$ and $M_{ge}$. This formation is discussed further in Ref.~\cite{Yao2023Energy}. Moreover, the clustering degree of these levels is quantified using the density of states $\rho_{\sigma}(E)$~\cite{note_DoS}, where variations in the width and height of the peak indicate the presence of quasi-degenerate and degenerate states, respectively. Notably, the $self$ pattern exhibits a more pronounced level clustering.

Despite the energy level clustering, each subband possesses a detailed internal structure, particularly noticeable at specific in-band positions, which could be described using multifractal energy spectra~\cite{Guarneri1993On,Jagannathan2021Fibonacci}. Instead of focusing solely on these spectra, the analysis utilizes $q$-weight scaling related to the wave function to examine the spatial differences between any two states. The spatial density of these CSs is illustrated using the scaled state density $Ar|\psi_{E_n}(i)|^2$, with the scaling factor $Ar$ corresponding to the lattice size.

Moreover, it is possibly stressed that a small energy-resolved window $\delta$ (approximately $10^{-3}t$) is used to depict the states within specific energy clusters accurately. This approach averages all states within the energy range $E_n\pm \delta/2$, highlighting the (quasi-)degenerate behavior in specific energy clusters. At the same time, it is used for subdiffusion behavior~\cite{De2020Subdiffusion}, such as benchmarking the autocorrelation function and mean-square displacement of the critical states in different lattices~\cite{OShaughnessy1985Diffusion,Guidoni1999Atomic,Basnarkov2006Diffusion,Di2022Diffusion}. These tools including the energy-correlation spectra~\cite{Yao2023Energy} are equivalent in capturing the energy spectra or states in statistical views. Here, we would analyze each state one by one.

\begin{figure*}[!htpb]
\centering
\includegraphics[width=6.35in]{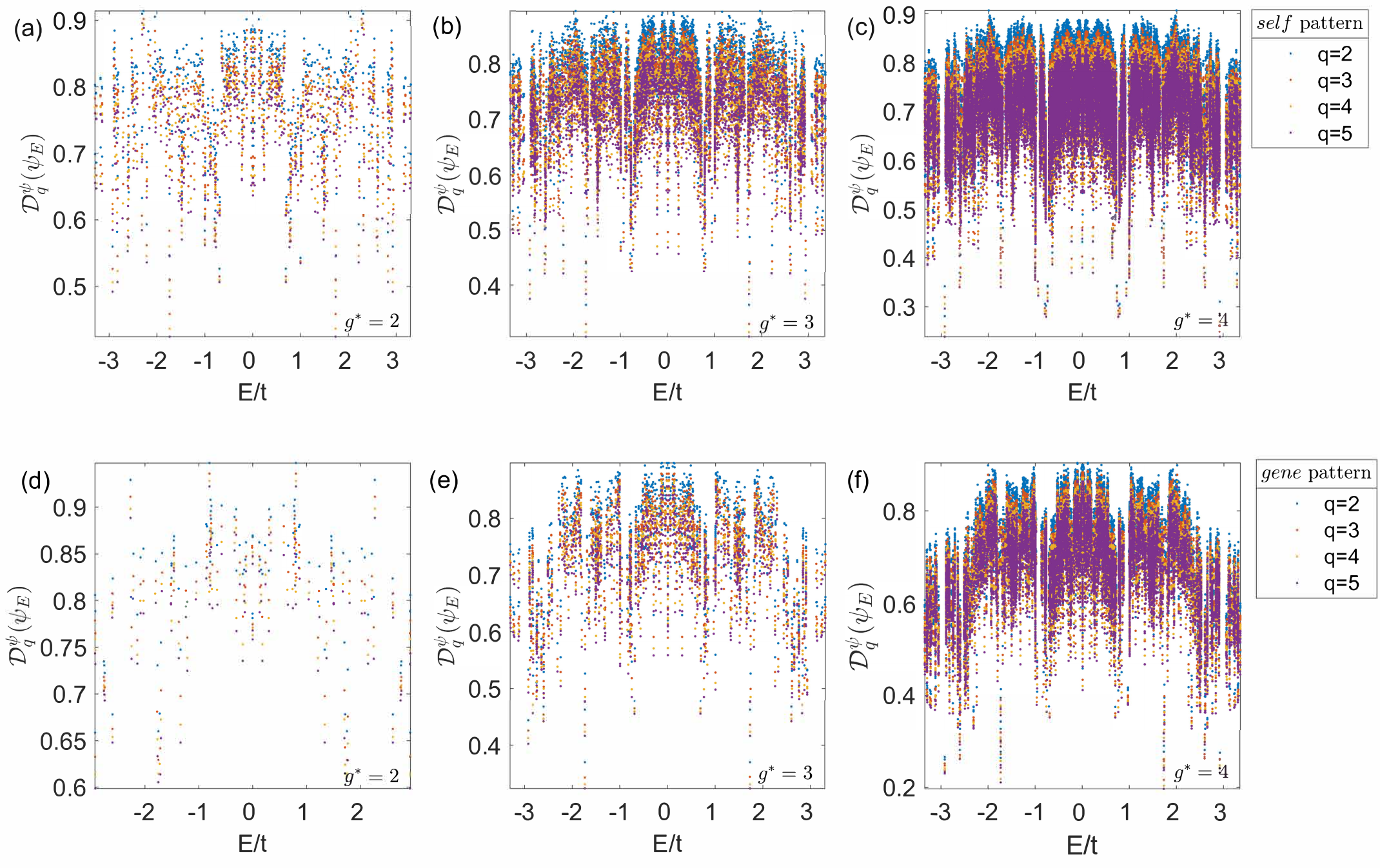}
\caption{(Color online) The $q$-order fractal dimension $\mathcal{D}_q^\psi(\psi_E)$ vs critical state $\psi_E(i)$ at allowed eigenenergy $E$. In order to comprise the impact of the dilation pattern, we still use the $generator(4,2)$ as a case. Considering the geometrical hierarchy effect in Ref~\cite{Yao2023Energy}, we modulate $g^*$ from 2 to 4: the $self$ pattern in (a), (b), and (c); and the $gene$ pattern in (d), (e), and (f). Here, $q$ starts from 2 to 5.}\label{Fig2:Multifra}
\end{figure*}

The second-order fractal dimension $\mathcal{D}_2^\psi(\psi_E)$ for all states in two $SC(4,2,$3$)$ lattices, associated with the $self$ and $gene$ patterns, typically ranges between 0.6 and 0.9. The proportion of states within this range is 0.974 and 0.937 for these two patterns, respectively. States outside this range are rare, indicating that the wave functions are predominantly critical. Fig.~\ref{Fig1:Sketch}(a) and Fig.~\ref{Fig1:Sketch}(b) illustrate four eigenstates with extreme values of $\mathcal{D}_2^\psi(\psi_E)$, showcasing the characteristic of partial occupancy in fractals and the spatial overlap between these states.

The eigenstates labeled $S1$, $S2$, $S3$, and $S4$ exhibit a semblance of approximate symmetry, likely resulting from spontaneous symmetry breaking. This characteristic remains consistent across various CSs and in different $SC(n,m,g^*)$ lattice configurations. However, these symmetrical traits in spatial wavefunction are not universally applicable to most CSs, as evidenced by $G4$ in Fig~\ref{Fig3:State}.

For the $vari$ pattern, the fractal dimension $\mathcal{D}_2^\psi(\psi_E)$ is around 0.3, as shown in Fig.~\ref{Fig1:Sketch}(c), which is notably low. This anomaly is primarily attributed to the substantial number of non-connecting site sub-clusters scattered throughout the entire lattice, which are dilated according to the $vari$ pattern. This structural feature reduces the connectivity and interference across the lattice, resulting in the significantly lower fractal dimension observed.

To make an analogy, we recall some electrical states in quasicrystals~\cite{stadnik2012physical}. First, taking the 1D Fibonacci chain~\cite{Jagannathan2021Fibonacci} case, some electronic states at the band center are self-similar and critical, having $\psi_m \propto(1 / N)^{\alpha_E}$ (where $\alpha_E$ is the exponent index~\cite{Kohmoto1987Critical,Fujiwara1989Multifractal}). Second, the situations become different in Penrose lattice (space dimension $D=2$), confined states~\cite{Tokihiro1988Exact,Fujiwara1989Electronic} and self-similar state~\cite{Tokihiro1988Exact} exist in special tiling alignments. Third, in Amman-Kramer lattice ($D=3$), the electronic states are also critical and have power-law decaying~\cite{Rieth1998Numerical}. $\mathcal{D}_q^\psi(\psi_E)$ for electronic states in quasicrystals is less than $D/2$, the value of $D$ depends on the space dimension where the quasicrystals are nested. And $\mathcal{D}_q^\psi(\psi_E)$ is less than 1 in \textit{Sierpi\'{n}ski} fractals, and it is closer to the fractal dimension $\mathcal{D}$. Additionally, the quasicrystals are lacuna-free, however, the fractals have lots of lacuna~\cite{Gefen1983Geometric,Lin1987Classi}. Hence the typical power-law character of some critical states is almost absent in fractals.

\subsection{The multifractality analysis in the $SC(n,m,g^*)$ lattices}~\label{SecIII:Multifrac}
For multifractal critical states, we could utilize the state-based fractal dimension $\mathcal{D}_q^\psi(\psi_E)$ to characterize their distinctive features, notably how the value of $\mathcal{D}_q^\psi(\psi_E)$  varies with the scaling parameter $q$~\cite{Hatsugai1990Energy,Cai2013Topological,Liu2015Localization,Wang2020Realization,Jagannathan2021Fibonacci,Xiao2021Observation}.
Without other statements throughout the article, the entire fractal lattice (where $Ar$ denotes the fractal lattice size) is considered, hence $\Omega=Ar$ is set. Generally, electronic states exhibit different behaviors when a single electron is situated at various energy levels, resulting in $\mathcal{D}_q^\psi(\psi_E)$ being dependent on the energy $E$, as depicted in Fig.~\ref{Fig2:Multifra}. The hole-particle symmetry further makes $\mathcal{D}_q^\psi(\psi_E)$ symmetric around $E=0$ across the entire spectrum of $\mathcal{D}_q^\psi(\psi_E)$.

\begin{figure}[!htp]
\centering
\includegraphics[width=3.2in]{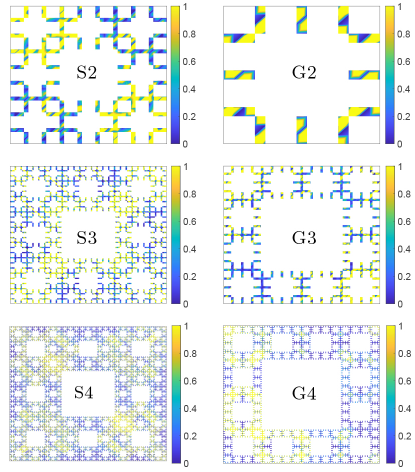}
\caption{(Color online) The scaled state density profile $Ar|\psi_E(i)|^2$ for six center states closest to the band center, are sequentially tagged as $S2$ in $SC(4,2,2)$ lattice, $S3$ in $SC(4,2,3)$ lattice, and $S4$ in $SC(4,2,4)$ lattice for the $self$ pattern; and $G2$ in $SC(4,2,2)$ lattice, $G3$ in $SC(4,2,3)$ lattice, and $G4$ in $SC(4,2,4)$ lattice for the $gene$ pattern. To visualize these center states vividly, the above six lattices are scaled in the same size. The scaled factor is the lattice size $Ar$, and the colorbar range maps from 0 to 1 (In units of $Ar$). }\label{Fig3:State}
\end{figure}
In the following, we would study how these CSs change with the geometrical hierarchy level $g^*$, the two dilation types of the $self$ and $gene$ pattern{\color{blue},} and the $generator(n,m)$.
First, we assess the influence of the geometric hierarchy level $g^*$, which is pivotal for self-similarity objects. In Fig.~\ref{Fig2:Multifra}(a), (b), and (c), we select the $generator(4,2)$ as a case. In the $self$ pattern, the $q$-weighted fractal dimension $\mathcal{D}_q^\psi(\psi)$ remains between 0.5 and 0.85 when $g^*=2$ in Fig.~\ref{Fig2:Multifra}(a); when $g^*$ increases to 3, $\mathcal{D}_q^\psi(\psi)$ undergoes slight adjustments but generally stays within the same range, as shown in Fig.~\ref{Fig2:Multifra}(b). Concurrently, many CSs cluster closely in the energy band and overlap within the spatial lattice, a phenomenon induced by the level attraction effect of $g^*$.  As $g^*$ escalates to 4, the maximum value we can simulate, the profiles of $\mathcal{D}_q^\psi(\psi)$ in Fig.~\ref{Fig2:Multifra}(c) resemble those at $g^*=3$.

We turn to the $gene$ pattern, when $g^*$ varies from 2 to 4, the entire span of $\mathcal{D}_q^\psi(\psi)$ narrows, and the distinct subclusters become apparent (refer to Fig.~\ref{Fig2:Multifra}(d), \ref{Fig2:Multifra}(e), and \ref{Fig2:Multifra}(f)). This effect is presumably due to the strong correlations between the CSs, whose energy levels are closely aligned, possibly leading to the anomalous level-spacing statistic $P(s)$~\cite{Yao2023Energy}.

We further fix the value of $g^*$ to compare the influence of dilation patterns on these critical states. With $g^*=2$, the spectra of $\mathcal{D}_q^\psi(\psi)$ display identical contours, as seen in Fig~\ref{Fig2:Multifra}(a) and Fig~\ref{Fig2:Multifra}(d), indicating that the $generator(4,2)$ plays a decisive role in shaping the electronic profile within fractal lattices. This behavior becomes more pronounced when increasing $g^*$ to 3 (Fig~\ref{Fig2:Multifra}(b) and Fig~\ref{Fig2:Multifra}(e)) or 4 (Fig~\ref{Fig2:Multifra}(c) and Fig~\ref{Fig2:Multifra}(f)). It is noteworthy that the overall fluctuation range of $\mathcal{D}_q^\psi(\psi)$ in the six scenarios is predominantly influenced by the lattice size $Ar$ of $SC(4,2,g^*=2\textbf{-}4)$, where $Ar$ is readily determined by the perimeter-area law~\cite{Mandelbrot1977fractals,feder2013fractals,Yao2023Energy}.

In the six scenarios mentioned above, we further analyze six approximated center-states in the energy band (specifically, $E\simeq 0$), labeled as $S2$-$S4$ and $G2$-$G4$ in Fig.~\ref{Fig3:State}). These multifractal critical states involve wave interference on a size scale relative to the entire lattice. Five of these states ($S2$-$S4$ and $G2$-$G3$) are nearly symmetrical, which might be attributed to the disrupted translational symmetry and the discrete scaling invariance. Moreover, it is essential to highlight that, due to their unique structure, unlike in \textit{Sierpi\'{n}ski gasket}, the diversity between spatial bulk state and the boundary state in these fractals is challenging, which has been observed by the Hall states~\cite{Iliasov2020Hall} in fractal.

\begin{figure}[!htp]
\includegraphics[width=3.25in]{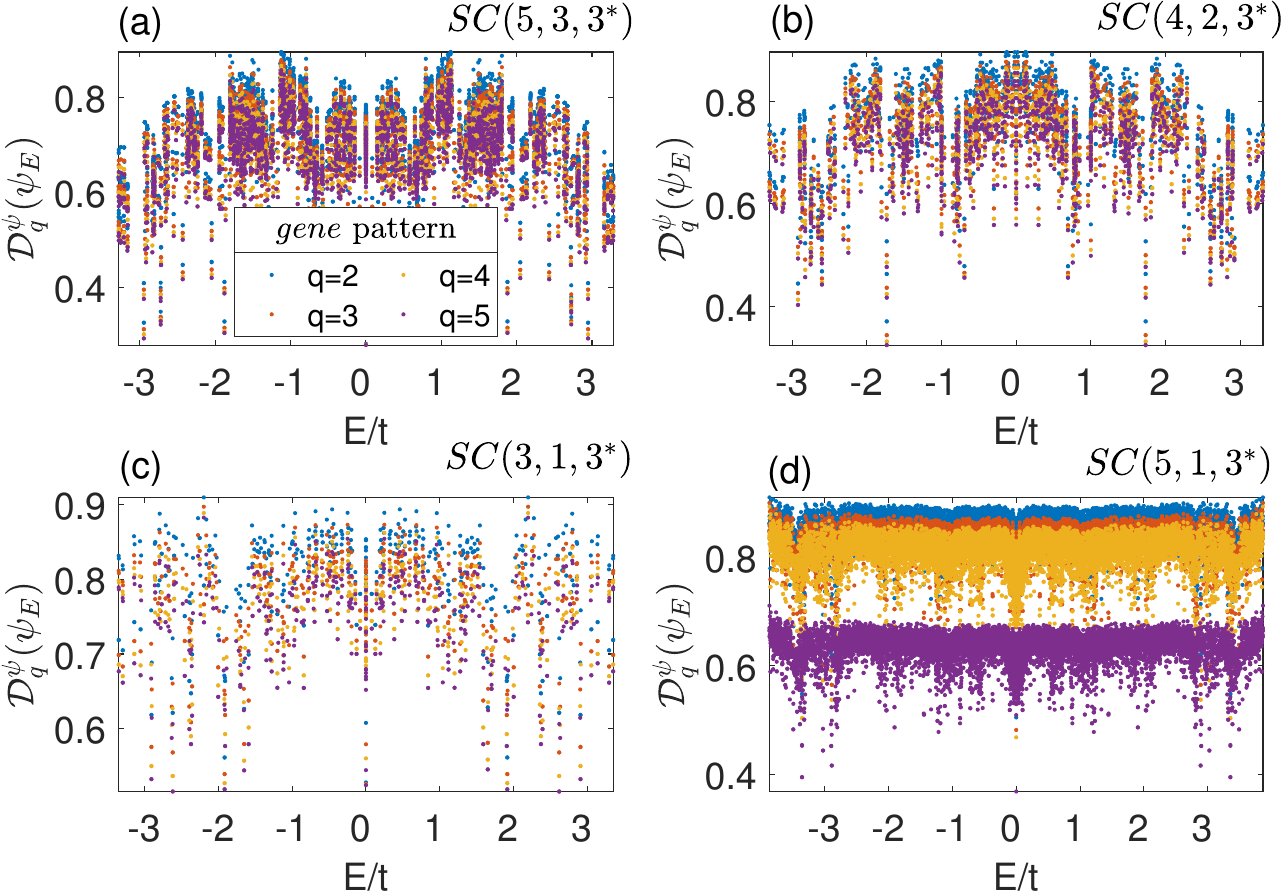}
\caption{(Color online) $\mathcal{D}_q^\psi(\psi_E)$ vs critical state $\psi_E(i)$ at allowed eigenenergy $E$ under the $gene$ pattern. The impact of the $generator(n,m)$ is comprised by taking $(5,3)$, $(4,2)$, $(3,1)$, and $(5,1)$ with the unchanged $g^*=3$, in four panels of (a), (b), (c), and (d). Here, $q$ is still from 2 to 5. Note that the averaged $\langle \mathcal{D}_2^\psi(\psi)\rangle$ in four cases is 0.7312, 0.7531, 0.8035, 0.8621, respectively.}\label{Fig3:generator}
\end{figure}

The \textit{Hausdoff dimension} $\mathcal{D}$ is essential for fractal objects and affects their quantum transport behaviors~\cite{van2016Quantum}. In our study, $\mathcal{D}_{se}$ is influenced not only by the choice of ($r$, $\mathcal{N}$), but also by the $generator(n,m)$ and geometrical hierarchy level $g^*$; $\mathcal{D}_{ge}$ is solely determined by its $generator(n,m)$. By applying the perimeter-area law~\cite{Mandelbrot1977fractals,feder2013fractals} to derive $\mathcal{D}$, $\mathcal{D}_{se}$ asymptotically follows $\mathcal{D}_{se} = \lg (\mathcal{N})/ \lg (r)$ in the $self$ pattern (considering a large value of $g^*$, $r$ and $\mathcal{N}$ refer in Ref.~\cite{Yao2023Energy}), and $\mathcal{D}_{ge}=\lg(n^{2}-m^{2})/\lg(n)$ in the $gene$ pattern~\cite{Yao2023Energy}. Note that $\mathcal{D}_{se}$ is only modulated in large fractal lattices, which makes simulation or experimental efforts difficult. We now wish to demonstrate the impact of $\mathcal{D}_{ge}$, and $\mathcal{D}_{se}$ is shown in Appendix~\ref{SS1}.

$\mathcal{D}_{ge}$ is generally varied by its $generator(n,m)$, making it easier to modulate in the case of small $g^*$ value. We select the sequence of $(5,3)$, $(4,2)$, $(3,1)$, and $(5,1)$, resulting in $\mathcal{D}_{ge}$ gradually increases from 1.7227 and 1.9746.
In Fig.~\ref{Fig3:generator}, we set $g^*=3$, its spectra of $\mathcal{D}_{ge}$ is significantly influenced by the $generator(n,m)$, as demonstrated in the four cases of (5,3), (4,2), (3,1) and (4,2). And We consider all the CSs, the averaged $\langle \mathcal{D}_2^\psi(\psi) \rangle$ for four cases is 0.7312, 0.7531, 0.8035, 0.8621, respectively. It is evident the increase of $\mathcal{D}_{ge}$ causes that the transition from the critical state to the extended state. There is a case in the $SC(5, 1, 3)$ lattice, as we discussed in Ref~\cite{Yao2023Energy}, for a $generator(n, m)$ with a large n and small m, the expanding lattices under the $gene$ pattern can be characterized as the translation-symmetry lattices with certain point-like or cluster-like defects. Consequently, their wave functions are slightly extended, and $\mathcal{D}_{ge}$ tends towards 1.
\section{Summary}\label{SecIV:Con.}
In summary, by scaling the inverse participation ratio and visualizing the wavefunction profile, we have analyzed electronic critical states upon fractal $SC(n, m, g^*)$ lattices that are percolated under two typical patterns of the $self$ pattern and the $gene$ pattern. We have found that the critical states generally exhibit the fragmentation behavior upon fractal $SC(n,m,g^*)$ lattices. Therefore, it causes the electronic states to behave multi-critically in fractals with $\mathcal{D}_q^\psi(\psi_E)$ lessing 1. Note that we ascribed the above confinement effects to the hierarchal properties of fractal structure. It could be observed by the autocorrelation function $C(t) \sim t^{-\gamma}$ and the mean-square displacement $d^2(t) \sim t^\delta$, with the predictions ($0< \gamma <1$ and $0< \delta <1$) for these critical states in more giant fractal lattices). For the $gene$ pattern and the $self$ pattern that we focus on, the fractal dimension $\mathcal{D}_2^\psi(\psi_E)$ is consistently greater than 0.5, which suggests that there is substantial spatial overlap among these critical states.

Besides, these states have the approximate symmetry when we pinned their energy near 0. As we know, physical states can be classified according to their transport properties. The Periodic Bloch functions describe the conducting states in crystalline systems, and the localized states in insulating systems exhibit exponential decaying. However, when discussing the critical states, the thing gets somewhat tricky. Generally speaking, they show strong spatial fluctuations at different scales~\cite{brezin2006applications}, occasionally accompanied by oscillatory behavior, which is evident in \textit{Sierpi\'{n}ski} fractals. We also emphasize that in aperiodic lattices, many mechanisms possibly prevent the wave function from decaying on large scales and not being constant over the entire lattice. There are some cases: i) for flux-threaded Koch fractals, under the action of commutation condition and special magnetic flux, $\psi(r_i)$ is certainly extended when pinning its eigenvalue in special energy window~\cite{Biswas2023Complete}; ii) for copper-mean chain~\cite{Sil1993Extended} or period-doubling chain~\cite{Chakrabarti1994Renormalization}, local cluster correlation makes $\psi(r_i)$ have the similar feature at some energy position, and some critical wave functions tend to expand in 1D Fibonacci chain~\cite{Macia1996Physical} and Thue-Morse lattice~\cite{Chakrabarti1995Role} due to short-range atom correlation; iii) both scale invariance and the finite order of ramification cause some states to be somewhat extended in Siperpinski gasket~\cite{Chakrabarti1996Exact,Biswas2023Designer}.

The interesting thing is that spectra of $\mathcal{D}_q^\psi(\psi_E)$ are determinedly modulated by the 'seed lattice' of $generator(n,m)$, and slightly changed with the geometrical hierarchy level $g^*$ and/or the dilation pattern. Additionally, the averaged fractal dimension $\mathcal{D}_2^\psi(\psi_E)$ would slightly raise with the $\mathcal{D}_{ge}$.

Note that comparing the observed properties in irregular objects that whose frontier is fractal-like, including the strong location traits in fractal drum~\cite{Even1999Localizations, Homolya2003Density}, the rich coherence of eigen wavefunction in Koch structure and Koch snowflake~\cite{Adrover2010Scaling}. Our work provides some insight about \textit{Sierpi\'{n}ski} lattice. the spectra of energy-level statistics~\cite{Yao2023Energy} and $\mathcal{D}_2^\psi(\psi_E)$ could assist the understanding the transport properties~\cite{van2016Quantum}, optical spectra~\cite{van2017Optical}, Hall conductivity~\cite{Iliasov2020Hall}, even superconductivity~\cite{iliasov2024strong}  in \textit{Sierpi\'{n}ski} lattices. At the same time, our work contributes to further study of disordered-induced localization or even many-body localization from the initial fractal-induced critical phase, and similar studies~\cite{Wang2020One, Zhou2023Exact} have been replicated in quasicrystals.

\section{Acknowledgments}
We thank Dr. Achille Mauri for discussing this work with us. This work was supported by the National Natural Science Foundation of China (Grant No. 12174291), the Natural Science Foundation of Hubei Province, China (2022BAA017, 2023BAA020), and the Core Facility of Wuhan University. Q.Y. acknowledges the China Scholarship Council (CSC) grant by file No. 202006270212 when attending Radboud University in the Netherlands.

\begin{appendix}
\setcounter{figure}{0}
\renewcommand{\thefigure}{S\arabic{figure}}
\vspace{10 pt}

\section{The spectra of $\mathcal{D}_q^\psi(\psi_E)$ in the $self$ pattern}~\label{SS1}
In the main text, our discussion is primarily focused on the $gene$ pattern. This pattern is crucial for understanding the behavior of single electron in fractal lattices, particularly in terms of the fractal dimensions and the localization properties of the wavefunctions.
\begin{figure}[!htbp]
\includegraphics[width=3.55in]{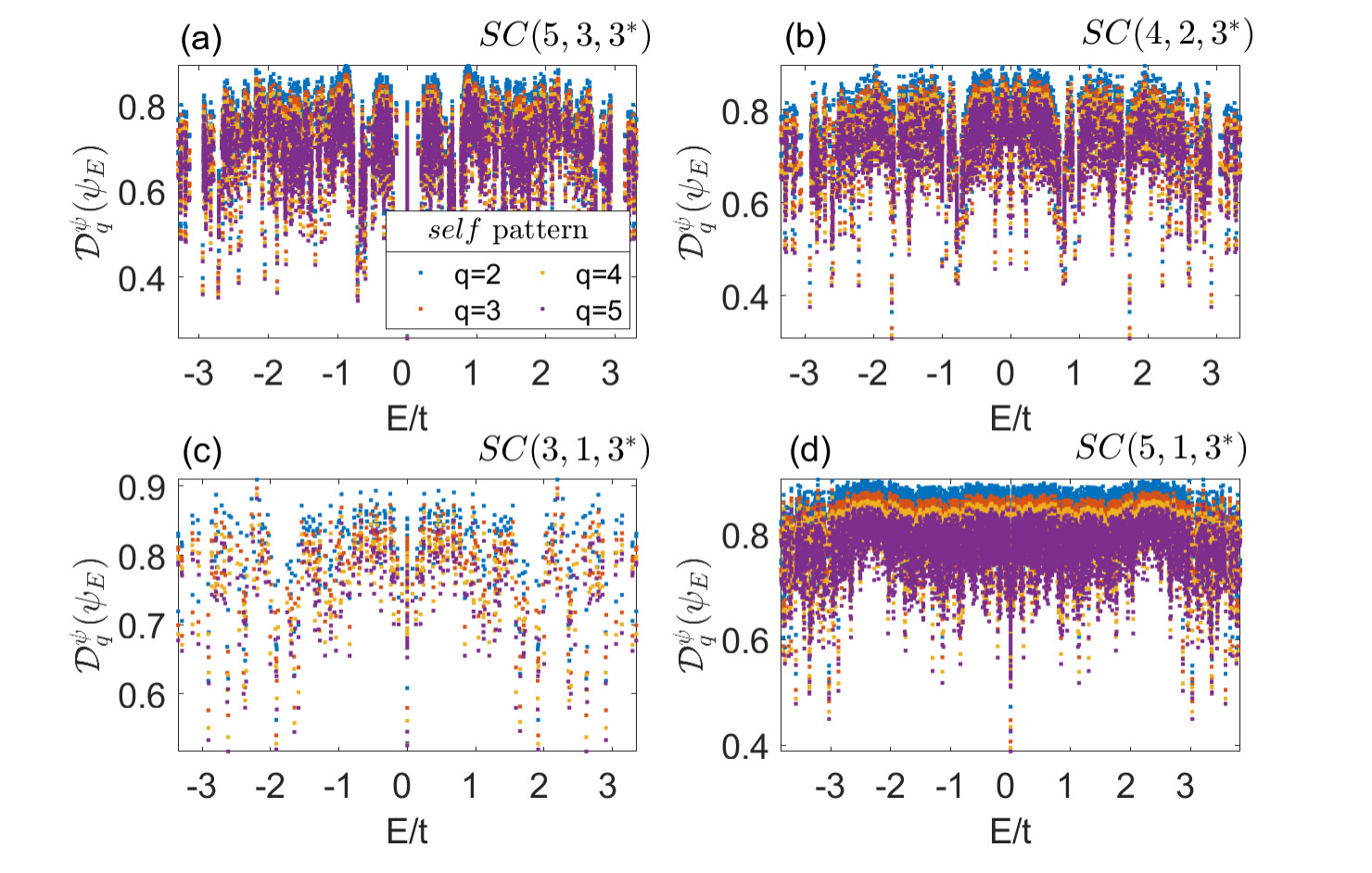}
\caption{(Color online) Similar to Fig.~\ref{Fig3:generator}, the impact of the $generator(n,m)$ is comprised under the $self$ pattern. And taking geometrical hierarchy level $g^*=3$, the $(n,m)$ of $(5,3)$, $(4,2)$, $(3,1)$, and $(5,1)$ are in (a), (b), (c), and (d), respectively. Here, $q$ is still from 2 to 5.}\label{Figs1:generator}
\end{figure}

To complement this analysis, we have included Fig.~\ref{Figs1:generator} with the $self$ pattern. This figure would illustrate how the spectra of $\mathcal{D}_q^\psi(\psi_E)$ correlates with the critical states $\psi_E(i)$ across different configurations. By fixing the geometrical hierarchy level $g^*$, we vary the $generator(n,m)$, and the overall sketch profile of $\mathcal{D}_q^\psi(\psi_E)$ is modulated substantially, see Fig.~\ref{Figs1:generator}(a), (b), (c), (d). Note that the $\mathcal{D}_q^\psi(\psi_E)$ also reaches 0.9 in Fig.~\ref{Figs1:generator}(d), which is due to the finite size effect. However, $\mathcal{D}_q^\psi(\psi_E)$ should decrease when taking the large value of $g^*$; it is beyond the possible simulation ability of our computation station.

\section{Constructing \textit{Sierpi\'{n}ski Carpet} $SC(n,m,g^*)$ lattices}~\label{SS2}
First, we need a ``seed" lattice (we named it the $generator(n,m)$, and $g^*$ is geometric hierarchy level) and dilation pattern including the $self$ pattern $M_{se}$ and the $gene$ pattern $M_{ge}$. Second, we construct the different types of $SC(n,m,g^*)$ lattice under the Eq.~\ref{es1},
\begin{eqnarray}\label{es1}
\textit{SC}(\mathrm{n}, \mathrm{m}, \mathrm{g})=M_{se,ge}(g) \otimes \textit{generator}(n, m),
\end{eqnarray}
where $g$ is hierarchy level, having $M_{se,ge}(g)\equiv M_{se,ge}(g-1)\otimes M_{se,ge}(1)$. And $g$ is distinguished from $g^*$, which is discussed in previous work~\cite{Yao2023Energy}. In Fig.~\ref{Fig1:Sketch}, $SC(4,2,3^*)$ lattices are constructed with the $generator(4,2)$ under three patterns (the third pattern is the $vari$ pattern, which is a variation of the $self$ pattern). Here, we only focus on the first two patterns.

\begin{figure}[!htbp]
\includegraphics[width=3.35in]{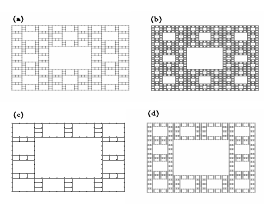}
\caption{(Color online) Four $SC(4,2,g^*)$ lattices in two dilation pattern. The same ``seed" lattice of the $generator(4,2)$ is used to construct the $SC(4,2,2)$ (a) and $SC(4,2,3)$ (b) in the $self$ pattern, and the $SC(4,2,2)$ (c) and $SC(4,2,3)$ (d) in the $gene$ pattern.}\label{Figs2}
\end{figure}

For the $self$ pattern, comprising the $\textit{SC}(n,m,g^*-1)$ lattice, the perimeter length of the $\textit{SC}(n,m,g^*)$ lattice at the $g^*$-th iteration increases by $r$ times and its area increases by $\mathcal{N}$ times. For consistency with Ref~\cite{Yao2023Energy}, supposing $M_{se}=[1,1,1; 1,0,1; 1,1,1]$ with $r=3$ and $\mathcal{N}=8$, see Fig.~\ref{Figs2}(a) and (b). For the $gene$ pattern, its pattern depends on its ``seed" lattice of the $generator(4,2)$,  and we have $M_{ge}(1)=[1,1,1,1; 1,0,0,1; 1,0,0,1; 1,1,1,1]$, see Fig.~\ref{Figs2}(c) and (d).

\end{appendix}


\begin{thebibliography}{71}
\bibitem [{Yao()}]{Yao2022talk}%
  \BibitemOpen
  \href@noop {} {}\bibinfo {note} {The fractals-induced critical phase could be
  driven to the localization phase by increasing the disorder strength, and one
  observe the energy level-correlation spectra gradually transitioning from the
  Wigner-like distribution to the Poisson distribution, and multifractal
  critical states trend to the localized states.}\BibitemShut {Stop}%
\bibitem [{not()}]{note_DoS}%
  \BibitemOpen
  \href@noop {} {}\bibinfo {note} {The stair-like DoS, $\rho(E)=1/Ar
  \sum_{k=1}^{Ar} \delta(E-E_k)$, would be blurred with a smooth function
  $g_\sigma (t)$. When we regularize the Dirac function $\delta(t)$ by Gaussian
  or Lorentz function $g_\sigma (t)$, $\rho_{\sigma}(E)$ usually has a smoother
  curve.}\BibitemShut {Stop}%
\end{thebibliography}
%
\end{document}